\documentclass[12pt]{article}

\usepackage{cite}
\usepackage{epsfig}
\usepackage{graphicx}
\usepackage{amsmath}
\usepackage{amssymb}
\usepackage{ulem}
\usepackage{mdwlist}          
\usepackage{color}              
\usepackage{hyperref}              

\usepackage{a41}
\usepackage{color}
\usepackage[rflt]{floatflt}
\usepackage{float}
\usepackage{slashed}

\usepackage{calrsfs}
\DeclareMathAlphabet{\pazocal}{OMS}{zplm}{m}{n}

\usepackage{array}

\usepackage[english]{babel}
\usepackage[latin1]{inputenc}
\usepackage[T1]{fontenc}
\usepackage{ae}

\usepackage{url}

\usepackage{amsmath, amsthm, amssymb}
\newtheorem{thm}{Theorem}[section]

\newtheorem{definition}[thm]{Definition}

 \newcommand{\GeV}{\mathrm{GeV}}
\newcommand{\HA}{{\rm H}}

\newcommand{\Mvec}{{\rm\bf M}}

\newcommand{\ep}{\varepsilon}

\usepackage{rotating}

\usepackage{graphicx}
\newcounter{mmacnt}
\def\restartmma{\setcounter{mmacnt}{0}}
\restartmma \catcode`|=\active
\def|#1|{\mathrm{#1}}
\catcode`|=12
\newenvironment{mma}{
 \par\smallskip
 \catcode`|=\active
 \parskip=0pt\parindent=0pt % locally
 \small
 \def\In##1\\{%
   \def\linebreak{\hfill\break\null\qquad}%
   \refstepcounter{mmacnt}
   \hangindent=2.5em\hangafter=0
   \leavevmode
   \llap{\tiny\sffamily In[\arabic{mmacnt}]:=\kern.5em}%
   \mathversion{bold}\footnotesize$\displaystyle##1$\normalsize
   \mathversion{normal}\par
 }%
 \def\\Print##1\\{%
   \def\linebreak{		\hfill\break}%
   \hangindent=2.5em\hangafter=0
   \leavevmode ##1\par}%
 \def\Out##1\\{%
   \def\linebreak{$\hfill\break\null\hfill$}%
   \kern\abovedisplayskip\par
   \hangindent=2.5em\hangafter=0
   \leavevmode
   \llap{\tiny\sffamily Out[\arabic{mmacnt}]=\kern.5em}
   \footnotesize$\displaystyle##1$\normalsize\hfill\null\par
   \kern\belowdisplayskip
 }%
 \def\Warning##1##2\\{%
   \def\linebreak{\hfill\break}%
   \hangindent=2.5em\hangafter=0
   \leavevmode
   {\scriptsize##1 : ##2}\par}%
}{%
 \par\smallskip
}

\usepackage{color}

\newenvironment{fshaded}{%
\MakeFramed {\FrameRestore}
}%
{\endMakeFramed}

\allowdisplaybreaks[4]

\begin{document}
\setlength{\baselineskip}{0.515cm}
\sloppy
\thispagestyle{empty}
\begin{flushleft}
DESY 25--185
%    \hfill {\tt arXiv:25xx.xxxxx [hep-ph]}
\\
CERN--TH--2025--259 \hfill December 2025
\\
MPP--2025--227
\\
RISC Report series 25--10
\\
\end{flushleft}

\mbox{}
\vspace*{\fill}
\begin{center}

{\Large\bf The heavy quark-antiquark asymmetry in the} 

\vspace*{2mm}
{\Large\bf variable flavor number scheme}

\vspace{3cm}
\large
A.~Behring$^a$, J.~Bl\"umlein$^{b,c}$, A.~De Freitas$^d$, A.~von~Manteuffel$^e$, \newline
C.~Schneider$^d$ and
K.~Sch\"onwald$^f$

\vspace{1.cm}
\normalsize
{\it $^a$~Max-Planck-Institut f\"ur Physik,
Boltzmannstra\ss{}e 8, 85748 Garching, Germany}

\vspace*{2mm}
{\it $^b$ Deutsches Elektronen-Synchrotron DESY, Platanenallee 6, 15738 Zeuthen, 
Germany} 

\vspace*{2mm}
{\it $^c$ Institut f\"ur Theoretische Physik III, IV, TU Dortmund, Otto-Hahn 
Stra\ss{}e 4, \newline 
44227 Dortmund, 
Germany}

\vspace*{2mm}
{\it $^d$~Johannes Kepler University,
Research Institute for Symbolic Computation (RISC),
Altenberger Stra\ss{}e 69,
                          A-4040, Linz, Austria}

\vspace*{2mm}
{\it $^e$~Institut f\"ur Theoretische Physik, Universit\"at Regensburg,
93040 Regensburg, Germany}

\vspace*{2mm}
{\it $^f$~CERN, Theoretical Physics Department, CH-1211 
Geneva 23, Switzerland}

\vspace*{3mm}

%%\today

\end{center}
\normalsize
\vspace{\fill}
\begin{abstract}
\noindent
The twist-2 heavy-quark and antiquark distributions, as defined in the variable flavor number 
scheme, turn out to be different due to QCD corrections from three-loop onward. This is caused 
by terms containing the color factor $d_{abc} d^{abc}$ in the heavy-flavor massive pure-singlet 
operator matrix elements (OMEs) $A^{\rm PS, s, (3)}_{Qq}$ for odd moments in the unpolarized case 
and for $\Delta A^{\rm PS, s, (3)}_{Qq}$ for even moments in the polarized case. The dependence on 
the factorization scale of the OMEs is ruled by the anomalous dimensions $\gamma^{\rm NS, s, (2)}_{qq}$ 
and $\Delta \gamma^{\rm NS, s, (2)}_{qq}$. The polarized calculations are performed in the Larin scheme. 
We compute the corresponding three-loop heavy-flavor distributions $(\Delta) f_Q(x,Q^2) 
- (\Delta) f_{\overline{Q}}(x,Q^2)$. Compared to the sum of the heavy-quark and antiquark parton 
distributions, 
their difference is small, however, non-vanishing. 
\end{abstract}

\vspace*{\fill}
\noindent
%\numberwithin{equation}{section}

\newpage

\vspace*{-1cm}
\noindent
%----------------------------------------------------------------------------------------------------------
\section{Introduction}
\label{sec:1}
%----------------------------------------------------------------------------------------------------------

\vspace*{1mm}
\noindent
Parton distributions rule a wide range of elementary particle phenomenology, and their
precise knowledge is instrumental for the study of many scattering processes 
Refs.~\cite{Accardi:2016ndt,Amoroso:2022eow}. In this context, a central question concerns the 
composition of the nucleons in terms of sea quarks and whether there are differences between
the sea quark and antiquark distributions.

The light-flavor quark and antiquark  distribution functions of the nucleons $u(x,Q^2)$, 
$d(x,Q^2)$, $s(x,Q^2), \overline{u}(x,Q^2), \overline{d}(x,Q^2), \overline{s}(x,Q^2)$ are 
of 
non-perturbative origin. Their first moments
%----------------------------------------------------------------------------------------------------------
\begin{eqnarray}
I_q = \int_0^1 dx [q(x,Q^2) - \overline{q}(x,Q^2)] 
\end{eqnarray}
%----------------------------------------------------------------------------------------------------------
obey the sum rules
%----------------------------------------------------------------------------------------------------------
\begin{eqnarray}
I_u = 2,~~~~
I_d = 1,~~~~
I_s = 0
\end{eqnarray}
%----------------------------------------------------------------------------------------------------------
for unpolarized protons. The sum rule for the strange quarks applies also to other 
higher mass pure sea quark species.
Here $x$ denotes the Bjorken variable, and $Q^2 = -q^2$ the 
virtuality in the deep-inelastic scattering process. In the polarized case, one has 
\cite{Blumlein:2010rn}\footnote{Here and in the following $\Delta$ marks quantities 
in the polarized  case.}
%----------------------------------------------------------------------------------------------------------
\begin{eqnarray}
I_{\Delta u} = ~0.928 \pm 0.014,~~~~
I_{\Delta d} = -0.342 \pm 0.018,
\end{eqnarray}
%----------------------------------------------------------------------------------------------------------
see also 
Refs.~\cite{Borsa:2024mss,Cruz-Martinez:2025ahf,Cocuzza:2025qvf}.
These constants are related to the hyperon $\beta$-decay parameters, cf.
Refs.~\cite{Lampe:1998eu,ParticleDataGroup:2008zun}.
While the up- and 
down-quark and antiquark distributions are different, and there is no $SU_F(3)$ sea 
quark symmetry \cite{NA51:1994xrz,PDG}, it has been discussed in 
Refs.~\cite{Signal:1987gz,Burkardt:1991di,Holtmann:1996be,Brodsky:1996hc,
Christiansen:1998dz,Cao:1999da,Melnitchouk:1999mv,Barone:1999di,Cao:2003ny,
Catani:2004nc,HERMES:2004zsh,Vega:2015hti,Zhu:2024tbu} that there is also a strange 
quark-antiquark difference. In Ref.~\cite{Catani:2004nc}, massless evolution effects from a starting 
scale $Q^2_0$ to a virtuality $Q^2$ were studied for strange, charm and bottom, concerning the creation of an asymmetry between quark and 
antiquark distributions, although, without considering mass effects.
In  Ref.~\cite{Brodsky:1996hc}, also a possible charm-anticharm 
difference in the intrinsic charm model~\cite{Brodsky:1980pb,Blumlein:2015qcn} was discussed. 
In the following, we consider only 
the so-called `extrinsic' contributions, which are calculated perturbatively in 
Quantum Chromodynamics (QCD) to three-loop order. 

Parton distributions at any twist \cite{Gross:1971wn} are no observables beyond lowest order 
in QCD \cite{Politzer:1974fr,Geyer:1977gv,Buras:1979yt,Reya:1979zk,Blumlein:2012bf}. As also 
the case for couplings and masses, one defines \'{e}talons in 
suitable schemes, as, e.g., the $\overline{\rm MS}$ scheme \cite{Bardeen:1978yd} or the Larin scheme \cite{Larin:1993tq},
to allow for comparisons. This also applies to the unpolarized and polarized twist-2 parton 
densities. 

The fixed flavor number scheme is based on describing the nucleon 
substructure by three massless parton distributions and the gluon distribution
at the level of twist-2 in deep-inelastic scattering.  Heavy-quark corrections
emerge as inclusive perturbative contributions from $O(a_s)$ onward, with $a_s = 
\alpha_s/(4\pi) = g_s^2/(16 \pi^2)$ the strong coupling constant, both in terms of real 
and virtual corrections. At very large virtualities $Q^2 \gg m_Q^2$, with $m_Q$ the heavy-quark mass, one 
may 
describe the heavy-flavor corrections to deep-inelastic scattering (DIS) in the 
variable flavor number 
scheme (VFNS) outlined in Ref.~\cite{Buza:1996wv},
by redefining the parton distributions. They now receive process-independent heavy-flavor corrections due 
to massive operator matrix elements. 
This is necessary to describe the massive Wilson coefficients in the asymptotic region $Q^2 \gg m_Q^2$
correctly, which is not possible in a pure massless approach. In this way, one also introduces 
the heavy-flavor parton distributions.

In the present paper, we calculate the heavy quark-antiquark asymmetry in the parton distributions
within the VFNS by exploiting computer algebra methods.
Flavor contributions of this kind
do not contribute to the well measured unpolarized and polarized structure functions 
$F_2(x,Q^2)$ 
and $g_1(x,Q^2)$, for which we derived the single-mass VFNS to three-loop order in 
Ref.~\cite{Ablinger:2025joi}.  
In the neutral current case,\footnote{The OMEs in the 
charged current case 
are different, as they also contain flavor excitation 
contributions, cf.~Refs.~\cite{Buza:1997mg,Blumlein:2014fqa}.} which we will consider in the 
following, heavy quark-antiquark difference terms
emerge in the $\gamma Z$-interference and $ZZ$ structure functions
$xF_3^{J_1,J_2}(x,Q^2)$ and 
$g_5^{J_1,J_2}(x,Q^2)$, with $J_k \in \{\gamma, Z\}$, 
cf.~Ref.~\cite{Blumlein:1996vs}.\footnote{One also could consider the structure function $g_4$, 
being related to $g_5$, cf.~Ref.~\cite{Blumlein:1996vs}.} 

The paper is organized as follows. In Section~\ref{sec:2}, we describe the basic formalism.
The unpolarized and polarized heavy quark-antiquark distribution asymmetries are calculated perturbatively 
in 
Section~\ref{sec:3}. Their logarithmic contributions due to the factorization scale are ruled by the anomalous 
dimensions~$(\Delta) \gamma_{qq}^{\rm NS, s, (2)}$, 
cf.~\cite{Moch:2004pa,Moch:2015usa,Blumlein:2021enk,Blumlein:2021ryt}. 
We have newly computed $\Delta \gamma_{qq}^{\rm NS, s, (2)}$ by using different methods.
In Section~\ref{sec:4}, 
we illustrate the flavor asymmetry for charm and bottom and compare to the sum of both distributions. 
Section~\ref{sec:5} contains 
the conclusions. 
We attach 
ancillary files of the OMEs in Mellin-$N$ and $x$-space, as well as a {\tt Fortran} code 
for their numerical evaluation.
%----------------------------------------------------------------------------------------------------------
\section{Basic Formalism}
\label{sec:2}
%----------------------------------------------------------------------------------------------------------

\vspace*{1mm}
\noindent
In the following we will work in Mellin-$N$ space, using the transformation 
%------------------------------------------------------------------------------------------
\begin{eqnarray}
\Mvec[f(x)](N) = \int_0^1 dx~x^{N-1} f(x)
\end{eqnarray}
%------------------------------------------------------------------------------------------
for the functions $f(x)$ given in momentum fraction $x$-space.
In the single-mass VFNS \cite{Buza:1996wv,Ablinger:2025joi}, the sum and difference of the heavy-quark 
contributions are given by the following relations
%------------------------------------------------------------------------------------------
\begin{eqnarray}
\label{eq:FQplus}
(\Delta) f_{Q + \overline{Q}} &\equiv& 
(\Delta) f_Q(N,Q^2,N_F+1) + (\Delta) f_{\overline{Q}}(N,Q^2,N_F+1) 
\nonumber\\
&=&
(\Delta) A_{Qq}^{\rm PS} \cdot (\Delta) \Sigma^+(N,Q^2,N_F)
+ (\Delta) A_{Qg} \cdot (\Delta) G(N,Q^2,N_F),
\\
\label{eq:FQminus}
(\Delta) f_{Q - \overline{Q}} &\equiv& 
(\Delta) f_Q(N,Q^2,N_F+1) - (\Delta) f_{\overline{Q}}(N,Q^2,N_F+1) 
\nonumber\\
&=& 
(\Delta) {A}^{\rm PS, s}_{Qq} \cdot (\Delta) \Sigma^-(N,Q^2,N_F).
\end{eqnarray}
%------------------------------------------------------------------------------------------
The massive OMEs $(\Delta) A_{Qq}^{\rm PS}$ and $(\Delta) A_{Qg}$ were computed to 
three-loop order
in Refs.~\cite{Ablinger:2014nga,Ablinger:2019etw,Ablinger:2023ahe,Ablinger:2024xtt}. The flavor combination
in Eq.~(\ref{eq:FQplus}) contributes to the heavy-flavor corrections to the structure functions $F_2$ and 
$g_1$, 
respectively. The OMEs $(\Delta) {A}^{\rm PS, s, (3)}_{Qq}$ are calculated in the present paper. 
For the quark contributions, the heavy-quark distributions are driven by the distributions
%------------------------------------------------------------------------------------------
\begin{eqnarray}
(\Delta) \Sigma^{\pm}  &=& [(\Delta) u \pm (\Delta) \overline{u}]
+ [(\Delta) d \pm (\Delta) \overline{d}] + [(\Delta) s \pm (\Delta) \overline{s}],
\end{eqnarray}
%------------------------------------------------------------------------------------------
and for the sum, also by the gluon distributions $(\Delta) G$.
The emergence of the color factor $d_{abc} d^{abc}$ in $(\Delta) {A}^{\rm PS, s, (3)}_{Qq}(N)$
is caused by the diagrammatic topology of $(\Delta) {A}^{\rm PS, (3)}_{Qq}(N)$ in the single-mass case, 
cf.~Refs.~\cite{Ablinger:2014nga,Ablinger:2019etw}, taking the odd moments for 
${A}^{\rm PS, (3)}_{Qq}$ and
the even moments for $\Delta {A}^{\rm PS, (3)}_{Qq}$. In the pure-singlet case,
the external lines are (directed) massless fermions. One could, as well, consider the
OME $(\Delta) {A}_{Qg}^{(3)}(N)$ with the same choice of moments. We checked that individual 
diagrams contain $d_{abc} d^{abc}$ terms, but they add up to zero due to the fact that 
gluon propagators have no direction. Therefore, there is no gluonic term in Eq.~(\ref{eq:FQminus}).
The color factor $d_{abc} d^{abc}$ is given by
$d_{abc} d^{abc} = {(N_c^2-1)(N_c^2-4)}/{N_c} = {40}/{3}$ 
%%%% MATAD: C1 = D33 * NA, C2 = D33 * NC mit NA=(NC^2-1)
%%%% D33 = 1/16/NC*(NC^2-1)(NC^2-4)
and $N_c = 3$ in the case of QCD.\footnote{For different conventions used in the literature,
see, however, Ref.~\cite{Behring:2014eya}, Eq.~(381), for remarks.}

There are also two other non-singlet distributions, $(\Delta) D_{3,(8)}(N,Q^2)$,
%------------------------------------------------------------------------------------------ 
\begin{eqnarray} 
(\Delta) D_3^{\pm} &=& \Delta (u \pm \overline{u}) - \Delta (d \pm \overline{d}), \\
(\Delta) D_8^{\pm} &=& \Delta (u \pm \overline{u}) + \Delta (d \pm \overline{d}) 
- 2 \Delta (s \pm \overline{s}). 
\end{eqnarray} 
%------------------------------------------------------------------------------------------
By decoupling of a heavy-quark $Q$ in the VFNS, the distributions $(\Delta) D_{3,8}^\pm$
are modified by 
%------------------------------------------------------------------------------------------
\begin{eqnarray}
(\Delta) D_{3,8}^\pm(N,Q^2,N_F+1) = (\Delta) A_{qq,Q}^{{\rm NS}}~\cdot~(\Delta) 
D_{3,8}^\pm(N,Q^2,N_F),
\end{eqnarray}
%------------------------------------------------------------------------------------------
see Refs.~\cite{Buza:1996wv,Ablinger:2025joi}. In the unpolarized $+ (-)$ cases the even (odd)
moments of $A_{qq,Q}^{{\rm NS}}$ are taken and in the polarized case the odd (even) moments. 
The OMEs $ A_{qq,Q}^{{\rm NS}}$ were calculated in 
Ref.~\cite{Ablinger:2014vwa}. However, they map between 
massless quark distributions only.

The flavor combinations $(\Delta) f_{Q - \overline{Q}}$ emerge in electroweak
structure functions, such as the neutral current unpolarized structure function $xF_3(x,Q^2)$
and polarized structure function $g_5(x,Q^2)$. Their crossing relations, 
cf.~Ref.~\cite{Blumlein:1996vs}, are in accordance with the respective choice of moments 
mentioned before.
In the unpolarized case, $xF_3$ can be measured from 
%------------------------------------------------------------------------------------------
\begin{eqnarray}
B^-(\lambda) &=& \frac{x Q^4}{4 \pi \alpha^2 Y_- \kappa_Z(Q^2)} \left[\frac{d\sigma^+(-\lambda)}{dx dQ^2} - 
\frac{d\sigma^-(+\lambda)}{dx dQ^2}\right] 
\nonumber\\ &=&
(a_e - \lambda v_e) xF_3^{\gamma Z}(x,Q^2)
+ \kappa_Z(Q^2) [2 v_e a_e + \lambda (v_e^2 + a_e^2)] xF_3^{ZZ}(x,Q^2),
\end{eqnarray}
%------------------------------------------------------------------------------------------
cf.~Refs.~\cite{Derman:1973sp,Blumlein:1987xk,Bardin:1996ch}. Analogous relations hold in 
the polarized case.
Here $Y_- = 1 - (1-y)^2$, $y = P.q/l.q$, 
$P$ is the proton momentum, 
$l$ the lepton momentum, 
and $\lambda$ denotes 
the degree of the longitudinal lepton beam polarization. The labels $\pm$ of the cross sections 
$\sigma$ refer to the charge of the 
incoming charged lepton. 
The weak couplings of the electron are $v_e = -1/2 + 2 
\sin^2 \theta_W, a_e = -1/2$, with $\theta_W$ the electroweak mixing angel, and $\kappa_Z(Q^2) = 
Q^2/(Q^2 +M_Z^2)/(4 \sin^2 \theta_W \cos^2 \theta_W)$, where $M_Z$ denotes the $Z$-boson mass.
First experimental results on $B^-$ were measured by BCDMS \cite{Argento:1983dj} and 
later at HERA \cite{H1:2015ubc}. Future measurements of this quantity can be carried 
out in a possible later stage at EIC\footnote{We thank E.~Aschenauer and W.~Melnitchouk 
for  remarks.}, which requires also polarized positron measurements \cite{Boer:2011fh,WALLY}. 
The measurement is planned also within the LHeC project 
\cite{LHeCStudyGroup:2012zhm,LHeC:2020van}.

Let us now turn to the calculation of the OMEs $(\Delta) \hat{{A}}^{\rm PS, s, (3)}_{Qq}(N)$ under the above choice of 
moments. 
The unrenormalized massive on-shell OMEs read
%------------------------------------------------------------------------------------------
\begin{eqnarray}
\left. %%\frac{2}{1 \pm (-1)^N} 
(\Delta) \hat{{A}}^{\rm PS, s, (3)}_{Qq}(N)\right|_{d_{abc} 
d^{abc}} = 
{a}_s^3 \left(\frac{m_Q^2}{\mu^2}\right)^{3 \ep/2} \left[
\frac{1}{3\ep} (\Delta) \hat{\gamma}_{qq}^{\rm NS, s,(2)}(N) + (\Delta) {a}_{Qq}^{\rm 
PS, s, (3)}(N)\right] + 
O(\ep),
\end{eqnarray}
%------------------------------------------------------------------------------------------
with $\mu$ the factorization scale and $ \hat{f}(N_F) = f(N_F+1) - f(N_F)$, see also the 
conventions in the regular pure-singlet 
case $(\Delta) A_{Qq}^{\rm PS}$ in Refs.~\cite{Ablinger:2014nga,Ablinger:2019etw}. Here the dimensional 
parameter is 
defined by $\ep = D - 4$, with $D$ the space-time dimension.

Because these OMEs start at $O(a_s^3)$, the only renormalization concerns the local
operator insertion
%------------------------------------------------------------------------------------------
\begin{eqnarray}
(\Delta) {A}^{\rm PS, s, (3)}_{Qq}(N) &=& Z_{qq}^{-1, \rm PS, s}
\left. %%\frac{2}{1 \pm (-1)^N} 
(\Delta) \hat{{A}}^{\rm PS, s, (3)}_{Qq}(N)\right|_{d_{abc} d^{abc}} 
\end{eqnarray}
%------------------------------------------------------------------------------------------
with
%------------------------------------------------------------------------------------------
\begin{eqnarray}
Z_{qq}^{-1, \rm PS, s} = 1 - a_s^3 \frac{1}{3 \varepsilon} (\Delta) 
\hat{\gamma}_{qq}^{\rm NS, s, (2)}(N).
\end{eqnarray}
%------------------------------------------------------------------------------------------
There is no mass nor coupling renormalization, and no collinear subtraction due to 
massless subsystems is needed, cf.~Ref.~\cite{Bierenbaum:2009mv}. The renormalized OME
is given by
%------------------------------------------------------------------------------------------
\begin{eqnarray}
\label{eq:DIFF}
(\Delta) {A}^{\rm PS, s, (3)}_{Qq}(N) &=& a_s^3 \left[\frac{1}{2} (\Delta) 
\hat{\gamma}_{qq}^{\rm NS, s, (2)}(N) 
\ln\left(\frac{m^2_Q}{\mu^2}\right) + (\Delta) {a}_{Qq}^{\rm PS, s, (3)}(N)\right],
\end{eqnarray}
%------------------------------------------------------------------------------------------
where $(\Delta) {a}_{Qq}^{\rm PS, s, (3)}$ denotes the constant part of the unrenormalized 
massive OME. All massive OMEs are solutions of 
renormalization group equations, see~Refs.~\cite{Buza:1996wv,Ablinger:2025joi}, due to 
which they account for scale evolution effects, which is also evident from their analytic 
structures in Mellin space, see Ref.~\cite{Bierenbaum:2009mv}.
Note that Eq.~(\ref{eq:DIFF}), derived in the VFNS, differs from Eqs.~(16, 
19) in a massless evolution approach in Ref.~\cite{Catani:2004nc}, especially by the 
non-logarithmic term $(\Delta) {a}_{Qq}^{\rm PS, s, (3)}$, not considered there, and the 
scale setting. In the present approach, the strange quark distribution is dealt with as a 
massless quark since $m_s < \Lambda_{\rm QCD}$, cf. Ref.~\cite{Agashe:2014kda}. 
%------------------------------------------------------------------------------------------
\section{The massive operator matrix elements}
\label{sec:3}
%------------------------------------------------------------------------------------------

\vspace*{1mm} 
\noindent
The technical steps of the present calculation are those described in previous 
papers, see, e.g.,~Ref.~\cite{Ablinger:2023ahe}. 
We use the packages 
{\tt QGRAF} \cite{Nogueira:1991ex}, {\tt Form}
\cite{Vermaseren:2000nd,Tentyukov:2007mu}, 
{\tt color} \cite{vanRitbergen:1998pn}, 
{\tt Reduze~2} \cite{Studerus:2009ye,vonManteuffel:2012np} for diagram generation,
the performance of the Lorentz- and Dirac algebra, color algebra, and the 
integration-by-parts reduction. The master integrals are calculated in 
Mellin $N$--space using different techniques, which are described in 
Refs.~\cite{Blumlein:2018cms,Blumlein:2022qci}. In the present case, 
only 
first--order--factorizable recurrences are obtained, which can be solved
by summation technologies based on difference ring theory
\cite{Karr:1981,Bron:00,Schneider:01,Schneider:04a,Schneider:05a,Schneider:05b,Schneider:07d,Schneider:2009rcr,
Schneider:10c,Schneider:15a,Schneider:08d,Schneider:08e,Schneider:2017,ABPET1}, 
encoded in the package
{\tt Sigma} \cite{SIG1,SIG2}.
The package {\tt HarmonicSums}
\cite{Vermaseren:1998uu,
Blumlein:1998if,
Ablinger:2013cf,
Ablinger:2011te,
Ablinger:2014bra,
Remiddi:1999ew,
Blumlein:2003gb,
Blumlein:2009ta,
Blumlein:2009cf,
Ablinger:2010kw,
Ablinger:2013hcp,
Ablinger:2014rba,
Ablinger:2015gdg,
ALL2016,
ALL2018,
Ablinger:2018cja,
Ablinger:2019mkx,
Ablinger:2021fnc}
is used to simplify the final expressions in Mellin-$N$ and $x$-space.
%------------------------------------------------------------------------------------------
\subsection{The operator matrix element \boldmath ${A}_{Qq}^{\rm PS, s, (3)}$}
\label{sec:31}
%------------------------------------------------------------------------------------------

\vspace*{1mm} 
\noindent
In the unpolarized case, one obtains the anomalous dimension 
\cite{Moch:2004pa,Blumlein:2021enk} 
%------------------------------------------------------------------------------------------
\begin{eqnarray}
\label{eq:anomDu}
\gamma_{qq}^{\rm NS, s, (2)} &=& 4 \frac{d_{abc} d^{abc}}{N_c} N_F \frac{1}{2}[1-(-1)^N]
\Biggl\{
        \frac{S_1 P_{13}}{(N-1) N^4 (1+N)^4 (2+N)}
\nonumber\\ &&        
        +\frac{2 P_{14}}{(N-1) N^5 (1+N)^5 (2+N)}
 +\Biggl[
                -\frac{2 P_{12}}{(N-1) N^3 (1+N)^3 (2+N)}
\nonumber\\ &&	       
                -\frac{4 \big(
                        2+N+N^2\big)^2 S_1}{(N-1) N^2 (1+N)^2 (2+N)}
        \Biggr] S_{-2}
 -\frac{\big(
                2+N+N^2\big)}{N^2 (1+N)^2} [S_3 - 2 S_{-3} + 4 S_{-2,1}]
\Biggr\}
\end{eqnarray}
%------------------------------------------------------------------------------------------
and the constant part of the unrenormalized OME in Mellin space
%------------------------------------------------------------------------------------------
\begin{eqnarray}
\lefteqn{{a}_{Qq}^{\rm PS, s, (3)} =} \nonumber\\ &&
\frac{4}{3} \frac{d_{abc} d^{abc}}{N_c}  \frac{1}{2}[1-(-1)^N]
\Biggl\{
        \frac{S_{2,1} P_1}{2 N^3 (1+N)^3 (2+N)}
        +\frac{S_1^2 P_3}{4 (N-1) N^4 (1+N)^4 (2+N)}
\nonumber\\ &&      
   +\frac{S_2 P_4}{4 (N-1) N^4 (1+N)^4 (2+N)}
        -\frac{3 \zeta_3 P_5}{2 (N-1) N^3 (1+N)^3 (2+N)}
\nonumber\\ &&   
      +\frac{S_{-3} P_6}{2 (N-1) N^3 (1+N)^3 (2+N)}
        +\frac{S_{-2,1} P_7}{(N-1) N^3 (1+N)^3 (2+N)}
\nonumber\\ &&    
        +\frac{S_3 P_8}{2 (N-1) N^3 (1+N)^3 (2+N)}
     +\frac{P_{11}}{(N-1) N^6 (1+N)^6 (2+N)^2}
     +\frac{2+N+N^2}{N^2 (1+N)^2}  
\nonumber\\ &&
\times \Biggl[
                \Biggl[
                        \frac{\big(
                                42+11 N+11 N^2\big) S_3}{2 (N-1) (2+N)}
                        +\frac{\big(
                                14-19 N-19 N^2\big) S_{-2,1}}{(N-1) (2+N)}
                        -\frac{3 \big(
                                10+7 N+7 N^2\big) \zeta_3}{2 (N-1) (2+N)}
                \Biggr]
\nonumber\\ && \times
 S_1
                +\frac{\big(
                        -18+13 N+13 N^2\big) S_{-3} S_1}{2 (N-1) (2+N)}
                -\frac{4 S_{-2} S_2}{(N-1) (2+N)}
                +\frac{3 \big(
                        6+N+N^2\big) S_4}{2 (N-1) (2+N)}
\nonumber\\ && 
                -\frac{1}{2} S_2^2
                +\frac{\big(
                        -2-5 N-5 N^2\big) S_{-2}^2}{(N-1) (2+N)}
                -\frac{12 S_{-4}}{(N-1) (2+N)}
                -\frac{3 \big(
                        14+N+N^2\big) S_{3,1}}{2 (N-1) (2+N)}
\nonumber\\ && 
      -\frac{6 \big(
                        -2+3 N+3 N^2\big) S_{-2,2}}{(N-1) (2+N)}
                -\frac{6 \big(
                        -2+3 N+3 N^2\big) S_{-3,1}
                }{(N-1) (2+N)}
\nonumber\\ &&                
 +\frac{12 \big(
                        -2+3 N+3 N^2\big) S_{-2,1,1}}{(N-1) (2+N)}
        \Biggr]
        +\Biggl[
                \frac{P_{10}}{4 (N-1) N^5 (1+N)^5 (2+N)^2}
\nonumber\\ &&              
  -\frac{(N-1) (2+N) \big(
                        1+2 N+2 N^2\big) S_2}{2 N^3 (1+N)^3}
        \Biggr] S_1
        -\frac{\big(
                2+N+N^2\big)^2 S_{-2} S_1^2}{(N-1) N^2 (1+N)^2 (2+N)}
\nonumber\\ &&        
+\Biggl[
                -\frac{2 S_1 P_2}{(N-1) N^3 (1+N)^3 (2+N)^2}
                +\frac{P_9}{2 (N-1) N^4 (1+N)^4 (2+N)^2}
        \Biggr] S_{-2}
\Biggr\},
\end{eqnarray}
%------------------------------------------------------------------------------------------
which is a new result. Here the nested finite harmonic sums are, 
cf.~Refs.~\cite{Vermaseren:1998uu,Blumlein:1998if},
%------------------------------------------------------------------------------------------
\begin{eqnarray}
S_{b,\vec{a}}(N) = \sum_{k=1}^N \frac{({\rm sign}(b))^k}{k^{|b|}} S_{\vec{a}}(k),~~
b, a_i \in \mathbb{Z} \backslash \{0\}, S_\emptyset = 1, 
\end{eqnarray}
%------------------------------------------------------------------------------------------
setting $S_{\vec{a}}(N) \equiv S_{\vec{a}}$.
The polynomials $P_i$ are
%------------------------------------------------------------------------------------------
\begin{eqnarray}
P_1&=&-6 N^6-26 N^5-38 N^4-7 N^3+17 N^2+8 N+4,\\
P_2&=&2 N^7+11 N^6+20 N^5+39 N^4+48 N^3+40 N^2+48 N+16,\\
P_3&=&-3 N^8-12 N^7-16 N^6-6 N^5-30 N^4-64 N^3-73 N^2-40 N-12,\\
P_4&=&-N^8-6 N^7-8 N^6+20 N^5+40 N^4+4 N^3-109 N^2-136 N-60,\\
P_5&=&N^8-N^7-13 N^6-4 N^5-N^4-43 N^3-67 N^2-44 N-20,\\
P_6&=&6 N^8+27 N^7+17 N^6-28 N^5-53 N^4-13 N^3+36 N^2-32 N-24,\\
P_7&=&6 N^8+27 N^7+61 N^6+24 N^5-N^4+31 N^3+4 N^2+32 N+8,\\
P_8&=&15 N^8+63 N^7+89 N^6+12 N^5-125 N^4-163 N^3-203 N^2-132 N-68,\\
P_9&=&-3 N^9-14 N^8-28 N^7+52 N^6+141 N^5+22 N^4-38 N^3+36 N^2+72 N+16,\\
P_{10}&=&-11 N^{11}-67 N^{10}-126 N^9+6 N^8+297 N^7-175 N^6-1582 N^5-2468 N^4
\nonumber\\ &&
-2358 N^3
-1492 N^2-616 N-112,\\
P_{11}&=&6 N^{12}+44 N^{11}+140 N^{10}+246 N^9+254 N^8+85 N^7+7 N^6+410 N^5
+873 N^4
\nonumber\\ &&
+861 N^3+478 N^2+156 N+24,
\\
P_{12}&=&N^6+3 N^5-8 N^4-21 N^3-23 N^2-12 N-4,\\
P_{13}&=&-3 N^8-12 N^7-16 N^6-6 N^5-30 N^4-64 N^3-73 N^2-40 N-12,\\
P_{14}&=&N^8+4 N^7+13 N^6+25 N^5+57 N^4+77 N^3+55 N^2+20 N+4.
\end{eqnarray}
%------------------------------------------------------------------------------------------
The first moment $N=1$ both of the anomalous dimension $\gamma_{qq}^{\rm NS, s, (2)}$ 
and  of ${A}_{Qq}^{\rm PS, (3),s}(N)$ vanish.
The expression in $x$-space, ${a}_{Qq}^{\rm PS, s, (3)}(x)$, 
is given in an ancillary file to this paper. 
It can be expressed by harmonic polylogarithms \cite{Remiddi:1999ew} up to weight {\sf  w 
= 5},
%------------------------------------------------------------------------------------------
\begin{eqnarray}
\HA_{b,\vec{a}}(x) &=& \int_0^x dy f_b(y) \HA_{\vec{a}}(y),~~b, a_i \in \{-1,0,1\},
\HA_\emptyset = 1,~~f_c(x) \in \left\{\frac{1}{1+x}, \frac{1}{x}, \frac{1}{1-x}\right\}
~{\rm with}
\nonumber\\
&& \HA_{\tiny \underbrace{0,...,0}_k}(x) :=  \frac{1}{k!} \ln^k(x). 
\end{eqnarray}
%------------------------------------------------------------------------------------------
In the small-$x$ region one obtains
%------------------------------------------------------------------------------------------
\begin{eqnarray}
\label{eq:NSUNP2}
{a}_{qq}^{\rm PS, s, (3)}(x) &\propto&
\frac{d_{abc} d^{abc}}{3 N_c}  \Biggl\{
        -4 \big(
                16
                +28 \zeta_3
                +13 \zeta_5
        \big)
        +\big(
                186
                -28 \zeta_3
        \big) \zeta_2
        -\frac{43}{5} \zeta_2^2
\nonumber\\ &&   
      +\big[
                84
                -4 \zeta_2
                -42 \zeta_2^2
                +4 \zeta_3
        \big] \ln(x)
        +\big[
                30
                +9 \zeta_2
                -28 \zeta_3
        \big] \ln^2(x)
        +\left[
                \frac{32}{3}
                -6 \zeta_2
        \right] \ln^3(x)
\nonumber\\ &&
        -\frac{1}{2} \ln^4(x)
        +\frac{1}{5} \ln^5(x)
\Biggr\},
\end{eqnarray}
%------------------------------------------------------------------------------------------
and for large $x$
%------------------------------------------------------------------------------------------
\begin{eqnarray}
{a}_{Qq}^{\rm PS, s, (3)}(x) &\propto&
\frac{d_{abc} d^{abc}}{3 N_c}  (1-x) \Biggl\{
        -20
        +13 \zeta_2
        -\frac{21}{5} \zeta_2^2
        +6 \zeta_3
        +\Biggl[
                17
                -8 \zeta_2
                -8 \zeta_3
        \Biggr] \ln(1-x)
\nonumber\\ &&
        + 
               [ -3
                +2 \zeta_2] \ln^2(1-x)
\Biggr\}.
\end{eqnarray}
%------------------------------------------------------------------------------------------

%--------------------------------------------------------------------------------------------------------------------------- 
\begin{figure}[H]
\centering
\includegraphics[width=0.60\textwidth]{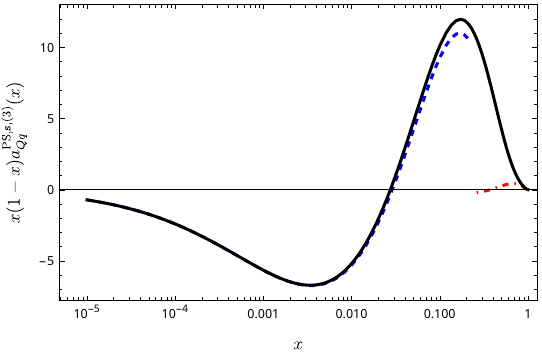}
\caption{\sf The constant part of the unrenormalized massive OME
$\hat{{A}}_{Qq}^{\rm PS, s, (3)}$, ${a}_{Qq}^{\rm PS, s, (3)}$, rescaled by $x(1-x)$. 
Dashed 
line: small-$x$ expansion up to 
the constant term. Dash-dotted line: large-$x$ approximation.
Full line: complete result. 
}
\label{fig:1}
\end{figure}
%--------------------------------------------------------------------------------------------------------------------------- 
In Figure~\ref{fig:1} we illustrate the constant part of the unrenormalized massive OME
$\hat{{A}}^{\rm PS, s, (3)}_{Qq}$, ${a}_{Qq}^{\rm PS, s, (3)}$, as a function of 
$x$.  
It is remarkable that the small-$x$ expansion, Eq.~(\ref{eq:NSUNP2}), holds up to 
relatively 
large values of $x$.

%------------------------------------------------------------------------------------------
\subsection{The operator matrix element \boldmath $\Delta {A}_{Qq}^{{\rm PS},s}$}
\label{sec:32}
%------------------------------------------------------------------------------------------

\vspace*{1mm} 
\noindent
Since in the contributing diagrams the two insertions of $\gamma_5$ are on different fermion 
lines, we employ the Larin scheme \cite{Larin:1993tq} for the calculation of $\Delta 
{A}_{Qq}^{{\rm PS},s}$. We use three different methods to compute the anomalous dimension 
$\Delta \gamma_{qq}^{\rm NS, s, (2)}$:~{\it i)} 
the unrenormalized on-shell OME $\Delta \hat{{A}}^{\rm PS, s, (3)}$ with massive fermions 
for even moments, {\it ii)} the unrenormalized massless off-shell OME 
$\Delta \hat{A}^{\rm PS, s, (3)}$ for even moments, and {\it iii)} 
the forward Compton amplitude for the $\gamma Z$-interference structure function $g_5$,
see Ref.~\cite{Blumlein:1996vs}. Here the projector of Eq.~(4.14) in Ref.~\cite{Bonino:2025bqa} 
has been used, which is structurally the same as the one in Eq.~(11) of 
Ref.~\cite{Behring:2019tus}.
We got the same result in all cases,\footnote{Our previous calculation 
used the forward Compton amplitude, erroneously with a 
different projector for the structure function $g_5$, Ref.~\cite{Blumlein:2021ryt}, Eqs.~(38, 39) 
and Ref.~\cite{Moch:2015usa}, p.~436. It has now been 
corrected leading to Eq.~(\ref{anomPOLC}). After our calculation  was finished, we 
found that in an independent calculation in Ref.~\cite{Zhu:2025gts}, using a 
SCET approach, the same result has been obtained, if one refers to the attachment {\tt 
dPSLarin.m} there.} 
%%However, Eq.~(4.37) in that reference deviates from our result by a term 
%%$\propto x \zeta_2$.} 
%----------------------------------------------------------------------------------------
\begin{eqnarray}
\label{anomPOLC}
\Delta \gamma_{qq}^{\rm NS, s, (2)} &=& 4 \frac{d_{abc} d^{abc}}{N_c} N_F \frac{1}{2} [1+(-1)^N]
\Biggl\{
        \frac{S_1 Q_4}{N^4 (1+N)^4}
        +\Biggl[
                -\frac{2 \big(
                        1+N+N^2
                \big)
\big(2+N+N^2\big)}{N^3 (1+N)^3}
\nonumber\\ &&                 -\frac{4 (N-1) (2+N)}{N^2 (1+N)^2} 
 S_1
        \Biggr] S_{-2}
        -\frac{\big(
                2+N+N^2\big)}{N^2 (1+N)^2} [S_3 - 2 S_{-3} + 4 S_{-2,1}]
\Biggr\}.
\end{eqnarray}
%------------------------------------------------------------------------------------------

\noindent
The agreement of the 
results of {\it i)} and {\it ii)} shows that potential `alien' operators, 
cf., e.g., Ref.~\cite{Matiounine:1998ky}, play no role in the present case. 
Additionally, obtaining the anomalous dimension from the forward Compton amplitude
requires a different projector than the one used in 
Refs.~\cite{Moch:2015usa,Blumlein:2021ryt}. At three-loop order the 
anomalous dimension $\Delta \gamma_{qq}^{\rm NS, s, (2)}$ 
is scheme invariant. It also obeys the Drell-Yan-Levy rescaling relation in $x$-space
%------------------------------------------------------------------------------------------
\begin{eqnarray}
\label{eq:DYL}
F(x) = - x {\sf Re}\left[F\left(\frac{1}{x}\right)\right],
\end{eqnarray}
%------------------------------------------------------------------------------------------
%--------------------------------------------------------------------------------------------------------------------------- 
\begin{figure}[H]
\centering
\includegraphics[width=0.60\textwidth]{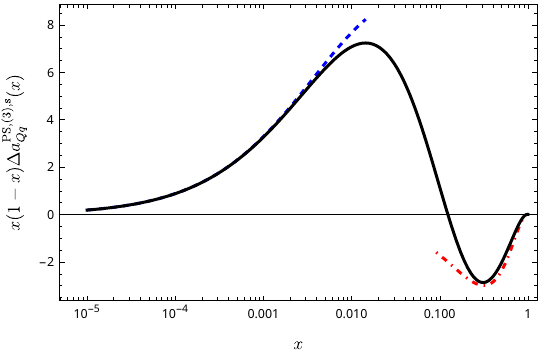}
\caption{\sf The constant part of the unrenormalized massive OME
$\Delta \hat{{A}}_{Qq}^{\rm PS, s, (3)}$, $\Delta {a}_{Qq}^{\rm PS, s, (3)}$, rescaled 
by 
$x(1-x)$. Dashed line: small-$x$ expansion up to the constant term. Dash-dotted line: 
large-$x$ approximation. Full line: complete result. 
}
\label{fig:2}
\end{figure}
%--------------------------------------------------------------------------------------------------------------------------- 

\vspace*{-1cm}

\noindent
see, e.g., Ref.~\cite{Blumlein:2000wh}, since it appears first at three-loop 
order.\footnote{JB 
thanks S.~Moch for reminding this relation.} 
Also the Mellin-inversion of Eq.~(\ref{eq:anomDu}) obeys Eq.~(\ref{eq:DYL}).

The projector given in Ref.~\cite{Behring:2019tus} was also applied to 
$\Delta \hat{{A}}^{\rm PS, (3)}_{Qq}$, i.e. the part $\propto [1 - (-1)^N]$,
from which the correct polarized 
three-loop anomalous dimensions $\Delta \gamma_{qq}^{\rm PS, (2)}$ was derived.
A corresponding projector, supplemented by a term $\propto p^2$, the off-shellness, 
needed to remove equation-of-motion terms, Eq.~(2.11) of 
Ref.~\cite{Blumlein:2021ryt},\footnote{See also Ref.~\cite{Blumlein:2022ndg}.} led  
to $\Delta \gamma_{qq}^{\rm PS, (2)}$ for the odd moments too. 

By method {\it i)} we also obtain the massive OME, $\Delta \hat{{A}}^{\rm PS, s, (3)}_{Qq}$, with
%------------------------------------------------------------------------------------------
\begin{eqnarray}
\lefteqn{\Delta {a}_{Qq}^{\rm PS, s, (3)}(N) =} \nonumber\\ &&
 \frac{4}{3} \frac{d_{abc} d^{abc}}{N_c} \frac{1}{2} [1+(-1)^N]
\Biggl\{
        \frac{S_{2,1} Q_2}{2 N^3 (1+N)^3}
        -\frac{3 \zeta_3 Q_3}{2 N^3 (1+N)^2}
        +\frac{S_1^2 Q_4}{4 N^4 (1+N)^4}
        +\frac{S_2 Q_5}{4 N^4 (1+N)^4}
\nonumber\\ && 
        +\frac{S_{-2,1} Q_8}{N^3 (1+N)^3}
        +\frac{S_3 Q_9}{2 N^3 (1+N)^3}
        +\Biggl[
                \frac{S_2 Q_1}{2 N^3 (1+N)^3}
                +\frac{Q_{10}}{2 N^5 (1+N)^5}
      -\frac{\big(
                        42-11 N-11 N^2\big) S_3}{2 N^2 (1+N)^2}
\nonumber\\ &&          
                +\frac{\big(
                        -14-19 N-19 N^2\big) S_{-2,1}}{N^2 (1+N)^2}
                -\frac{3 \big(
                        -10+7 N+7 N^2\big) \zeta_3}{2 N^2 (1+N)^2}
        \Biggr] S_1
        +\frac{\big(
                -2-N-N^2\big) S_2^2}{2 N^2 (1+N)^2}
\nonumber\\ &&
        +\frac{3 (N-2) (3+N) S_4}{2 N^2 (1+N)^2}
        +\Biggl[
                \frac{Q_6}{N^4 (1+N)^4}
                +\frac{2 \big(
                        4+12 N-3 N^3-N^4\big) S_1}{N^3 (1+N)^3}
                -\frac{(N-1) (2+N) S_1^2}{N^2 (1+N)^2}
\nonumber\\ &&               
 +\frac{4 S_2}{N^2 (1+N)^2}
        \Biggr] S_{-2}
        +\frac{\big(
                2-5 N-5 N^2\big) S_{-2}^2}{N^2 (1+N)^2}
        +\Biggl[
                \frac{\big(
                        18+13 N+13 N^2\big) S_1}{2 N^2 (1+N)^2}
\nonumber\\ &&              
 + \frac{Q_7}{2 N^3 (1+N)^3}
        \Biggr] S_{-3}
        -\frac{3 \big(
                -14+N+N^2\big) S_{3,1}}{2 N^2 (1+N)^2}
        -\frac{6 \big(
                2+3 N+3 N^2\big) S_{-2,2}
        }{N^2 (1+N)^2}
\nonumber\\ &&
        +\frac{12 S_{-4}}{N^2 (1+N)^2}
        -\frac{6 \big(
                2+3 N+3 N^2\big) S_{-3,1}}{N^2 (1+N)^2}
        +\frac{12 \big(
                2+3 N+3 N^2\big) S_{-2,1,1}}{N^2 (1+N)^2}
\Biggr\},
\end{eqnarray}
%-------------------------------------------------------------------------------------------
and the polynomials
%-------------------------------------------------------------------------------------------
\begin{eqnarray}
Q_1&=&-2 N^4-4 N^3-5 N^2-3 N-2, \\
Q_2&=&-6 N^5-20 N^4-10 N^3+N^2-3 N-2, \\
Q_3&=&N^5+5 N^4-8 N^3-3 N^2+3 N+6, \\
Q_4&=&-3 N^6-9 N^5-5 N^4+5 N^3+19 N^2+15 N+6, \\
Q_5&=&-N^6-N^5+7 N^4+7 N^3+19 N^2+15 N+6, \\
Q_6&=&N^6+7 N^5+25 N^4+12 N^3-20 N^2-31 N-10, \\
Q_7&=&6 N^6+15 N^5-24 N^4-52 N^3-39 N^2+6 N-4, \\
Q_8&=&6 N^6+15 N^5+24 N^4+12 N^3+N^2-18 N-4, \\
Q_9&=&15 N^6+60 N^5+42 N^4-45 N^3-37 N^2+3 N+6, \\
Q_{10}&=&-16 N^8-65 N^7-71 N^6+25 N^5+58 N^4+80 N^3+110 N^2+81 N+18.
\end{eqnarray}
%------------------------------------------------------------------------------------------
The expression in $x$-space is given in an ancillary file.
Here the first moment is non-vanishing. In the small-$x$ region one obtains 
%------------------------------------------------------------------------------------------ 
\begin{eqnarray} 
\label{eq:NSUNP2x} 
\Delta {a}_{Qq}^{\rm PS, s, (3)}(x) \propto 
\frac{1}{3} \frac{d_{abc} 
d^{abc}}{N_c}  \zeta_2 \ln(x) [(50 + 2 \zeta_2) - 9 \ln(x) - 6 \ln^2(x)] 
\end{eqnarray} 
%------------------------------------------------------------------------------------------ 
and for large-$x$
%------------------------------------------------------------------------------------------ 
\begin{eqnarray} 
\Delta {a}_{Qq}^{\rm PS, s, (3)}(x) \propto 
\frac{1}{3} \frac{d_{abc} 
d^{abc}}{N_c}  (1-x) 
        \big\{\big[
                14
                -4 \zeta_2
                -8 \zeta_3
        \big] \log (1-x)
        +\big[
                -3
                +2 \zeta_2
        \big] \log ^2(1-x)
\big\}.
\end{eqnarray} 
%------------------------------------------------------------------------------------------ 

\noindent
With the OMEs calculated in this paper, the set of massive single-mass OMEs at three-loop order 
is now complete,  extending the results reported in Refs.~\cite{Ablinger:2010ty,
Ablinger:2014lka,Ablinger:2014vwa,Ablinger:2014nga,Behring:2014eya,Ablinger:2022wbb,
Ablinger:2023ahe,Ablinger:2024xtt,Ablinger:2019etw,Blumlein:2021xlc,Behring:2021asx}.

%------------------------------------------------------------------------------------------
\section{The heavy quark-antiquark asymmetry}
\label{sec:4}
%------------------------------------------------------------------------------------------

\vspace*{1mm} 
\noindent
Finally, we calculate the heavy quark-antiquark difference and sum distributions,
$x[ f_Q(x,Q^2) \mp f_{\overline{Q}}(x,Q^2)]$ and $[(\Delta) f_Q(x,Q^2) \mp 
(\Delta) 
f_{\overline{Q}}(x,Q^2)]$ by setting $\mu^2 = Q^2$,
in the VFNS, for $Q = c, b$. In the unpolarized case, we refer 
to the parton distribution functions Ref.~\cite{Alekhin:2017kpj} from \cite{Buckley:2014ana}, and 
in the polarized case to those of Ref.~\cite{Blumlein:2024euz}. 

\begin{figure}[H]
\centering
\includegraphics[width=0.50\textwidth]{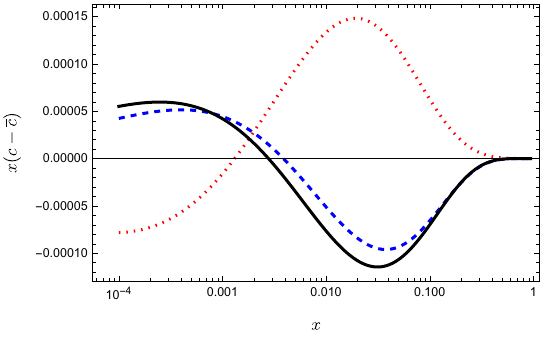}
\includegraphics[width=0.47\textwidth]{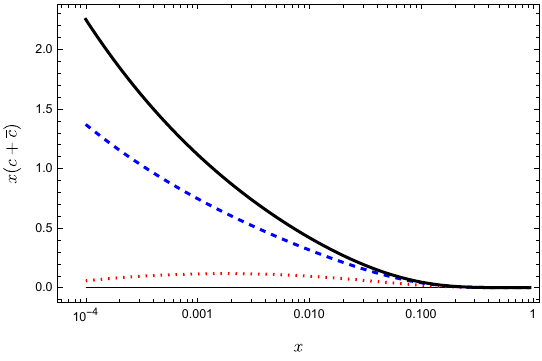}
\caption{\sf The unpolarized  distributions $x[c(x,Q^2)-\overline{c}(x,Q^2)]$ (left 
panel) 
and $x[c(x,Q^2)+\overline{c}(x,Q^2)]$ (right panel). Dotted lines: $Q^2 = 4~\GeV^2$.
Dashed lines: $Q^2 = 30~\GeV^2$. Full lines: $Q^2 = 100~\GeV^2$.} 
\label{fig:3}
\end{figure}

\begin{figure}[H]
\centering
\includegraphics[width=0.50\textwidth]{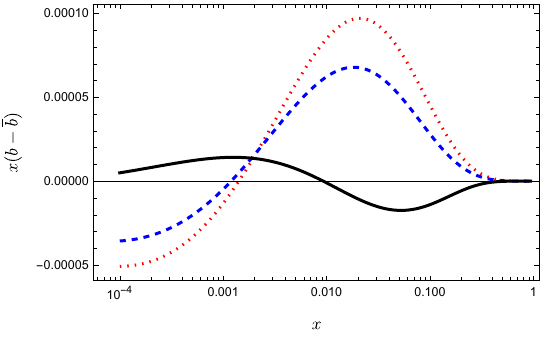}
\includegraphics[width=0.47\textwidth]{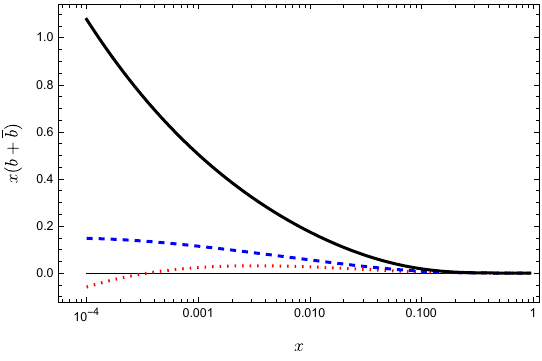}
\caption{\sf The unpolarized  distributions $x[b(x,Q^2)-\overline{b}(x,Q^2)]$ (left 
panel) 
and $x[b(x,Q^2)+\overline{b}(x,Q^2)]$ (right panel). Dotted lines: $Q^2 = m_b^2$.
Dashed lines: $Q^2 = 30~\GeV^2$. Full lines: $Q^2 = 100~\GeV^2$.} 
\label{fig:4}
\end{figure}

\begin{figure}[H]
\centering
\includegraphics[width=0.49\textwidth]{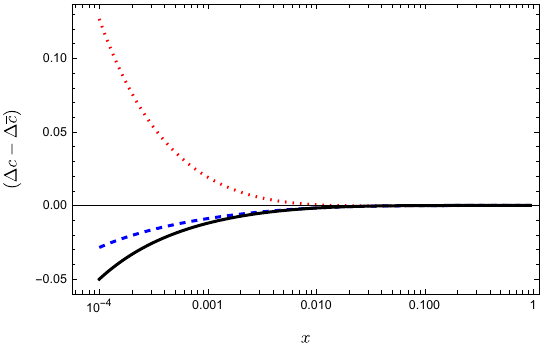}
\includegraphics[width=0.49\textwidth]{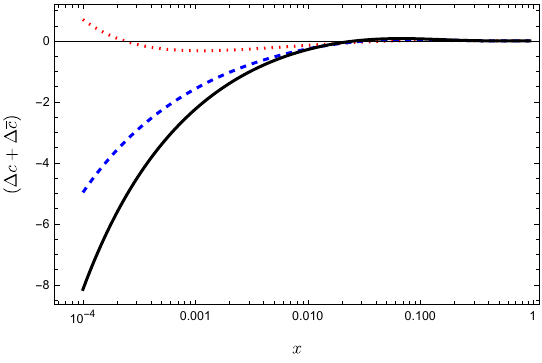}
\caption{\sf The polarized  distributions $[\Delta c(x,Q^2)-\Delta \overline{c}(x,Q^2)]$ 
(left panel) 
and $[\Delta c(x,Q^2)+ \Delta \overline{c}(x,Q^2)]$ (right panel). Dotted lines: $Q^2 = 
4~\GeV^2$.
Dashed lines: $Q^2 = 30~\GeV^2$. Full lines: $Q^2 = 100~\GeV^2$.}
\label{fig:5}
\end{figure}

\begin{figure}[H]
\centering
\includegraphics[width=0.49\textwidth]{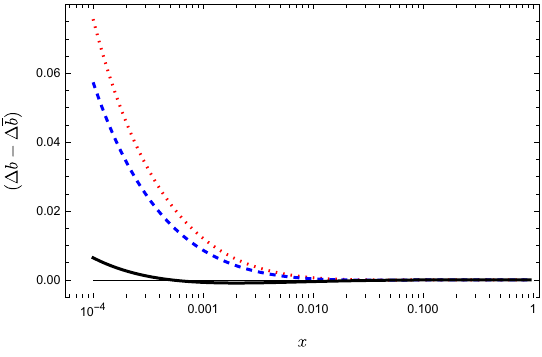}
\includegraphics[width=0.49\textwidth]{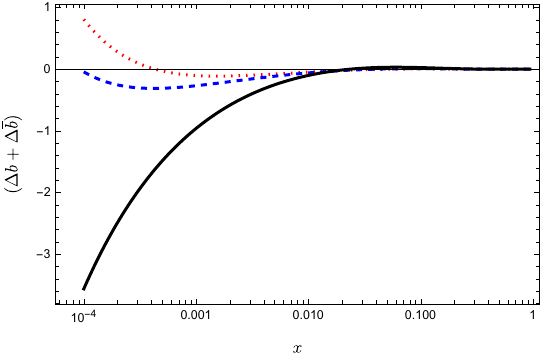}
\caption{\sf The polarized  distributions $[\Delta b(x,Q^2)-\Delta \overline{b}(x,Q^2)]$ 
(left panel) 
and $[\Delta b(x,Q^2)+ \Delta \overline{b}(x,Q^2)]$ (right panel). Dotted lines: $Q^2 = 
m_b^2$.
Dashed lines: $Q^2 = 30~\GeV^2$. Full lines: $Q^2 = 100~\GeV^2$.}
\label{fig:6}
\end{figure}

For the distributions shown in Figures~\ref{fig:3}--\ref{fig:6}, we refer to three 
massless flavors representing 
$\Sigma^\pm$ both for the charm and bottom distributions, Eq.~(\ref{eq:FQplus}, 
\ref{eq:FQminus}), which only differ by the logarithmic terms in the OMEs at 
the respective values of $Q^2$.
The heavy-quark masses in the on-shell scheme, used in the calculation of the massive OMEs,  
are \cite{Alekhin:2012vu,Agashe:2014kda}
%------------------------------------------------------------------------------------------
\begin{eqnarray}
m_c = 1.59~\GeV,~~~~~~m_b = 4.78~\GeV.
\end{eqnarray}
%------------------------------------------------------------------------------------------
The values of the strong coupling constant 
%%%%$\alpha_s(m_c^2) = 0.29385$,
$\alpha_s(4~\GeV^2) = 0.26897$,
$\alpha_s(m_b^2) = 0.20452$,
$\alpha_s(30~\GeV^2) = 0.1972$,
$\alpha_s(100~\GeV^2) = 0.1706$ are consistent with the value $\alpha_s(M_Z^2) = 0.1147$.
The {\tt Fortran} programs were designed by applying code optimization
\cite{Ruijl:2017dtg} and we use the numerical representation of harmonic polylogarithms 
up to {\sf w = 5} of Ref.~\cite{Gehrmann:2001pz}. Convolution integrals are calculated
by the package {\tt DAIND}, cf.~Ref.~\cite{AIND}.
 
%--------------------------------------------------------------------------------------------------------------------------- 
%--------------------------------------------------------------------------------------------------------------------------- 

In Figures~\ref{fig:3} to  \ref{fig:6}, we illustrate both the difference and the sum 
of the charm and bottom distributions, respectively, as functions of $x$ and $Q^2$. Note 
that in the unpolarized case, the OMEs $A_{Qq}^{\rm PS, (3)}$ and $A_{Qg}^{\rm (3)}$ have 
different signs, leading to partial cancellations. At the higher scales shown, the charm 
quark distributions are about twice bigger than those for the bottom quarks, see 
Figures~\ref{fig:3} and  \ref{fig:4}. The difference distributions $x[Q(x,Q^2) - 
\overline{Q}(x,Q^2)]$ take values in the range $-0.0001$ to $+0.0015$, which are 
oscillating since 
their first moments vanish. 

%--------------------------------------------------------------------------------------------------------------------------- 
%--------------------------------------------------------------------------------------------------------------------------- 

\vspace*{-1cm}
%--------------------------------------------------------------------------------------------------------------------------- 
%--------------------------------------------------------------------------------------------------------------------------- 
%--------------------------------------------------------------------------------------------------------------------------- 
%--------------------------------------------------------------------------------------------------------------------------- 

In the polarized case, we illustrate the quark-antiquark difference distributions 
for the number densities in Figures~\ref{fig:5} and \ref{fig:6}. 
Also here the charm quark distributions are about twice as 
large as those for bottom in the kinematic range shown, taking values between $-0.05$ 
and $0.12$, more peaked towards smaller values of $x$. Their measurement is even more 
difficult, as two polarization asymmetries have to be formed. The sum distributions are
widely negative in the small-$x$ region. Correspondingly, the 
contributions to the nucleon momentum and spin budget by the 
PDF-asymmetries are very small in the heavy-quark case.

In measuring $B^-(\lambda)$ off deuteron targets, both the 
distributions $D_8^-$ and $\Sigma^-$ contribute in the combination $xF_3^{\gamma Z} = 
1.39~xD_8^- + 2.44~x\Sigma^-$, and analogously in the polarized case. 
It turns out that in the VFNS the heavy quark-antiquark asymmetry 
$(\Delta) f_{Q - \overline{Q}}(x,Q^2)$ is very small but non-vanishing.
An experimental measurement is challenging and will require very large luminosities
and precision, despite of the fact that the heavy-flavor contributions 
at three-loop order are solely determined by the heavy-quark tagging part.
%---------------------------------------------------------------------------------------------$
\section{Conclusions}
\label{sec:5}
%---------------------------------------------------------------------------------------------$

\vspace*{1mm}
\noindent
We calculated the massive OMEs describing the perturbative creation of the asymmetry of the 
heavy-quark PDFs $(\Delta) f_Q(x,Q^2) - (\Delta) f_{\overline{Q}}(x,Q^2)$ in the unpolarized and polarized cases in 
QCD 
in the variable flavor number scheme. Unlike the sum
of the heavy-quark PDFs, which contribute from $O(a_s)$, their asymmetry occurs first at 
$O(a_s^3)$ in the VFNS. While the sum is driven by the PDFs $\Sigma^+$ and $G$, the 
difference results from $\Sigma^-$. The difference distributions 
contribute to the  polarization asymmetry $(\Delta) B^-(\lambda)$, measured by using 
polarized electron and 
positron deep-inelastic data. It turns out that the 
distributions $(\Delta) f_Q(x,Q^2) - (\Delta)f_{\overline{Q}}(x,Q^2)$ are non-vanishing but very small 
and
require huge luminosities to be measured. They contribute with a correspondingly 
small rate both to the nucleon momentum and nucleon spin. In the heavy-quark case,  
the quark and antiquark distributions are different in the VFNS.

We corrected the result for the polarized anomalous dimension $\Delta \gamma_{qq}^{\rm NS, s, (2)}$ 
in Refs.~\cite{Blumlein:2021ryt}, which has been calculated by us by three different methods.

\vspace*{3mm}
\noindent
{\bf Acknowledgment.} We thank J.~Ablinger, M.~Diehl, P.~Marquard, P.~Ploessl, and G.~Salam
for discussions. This work has been funded by the Austrian Science Fund (FWF) Grant
DOI 10.55776/P20347. KS is supported by the European Union under the HORIZON program in 
Marie Sk\l{}odowska-Curie project No. 101204018
\hspace*{-2cm} \parbox{180pt}{\centering\includegraphics[height=5mm]
{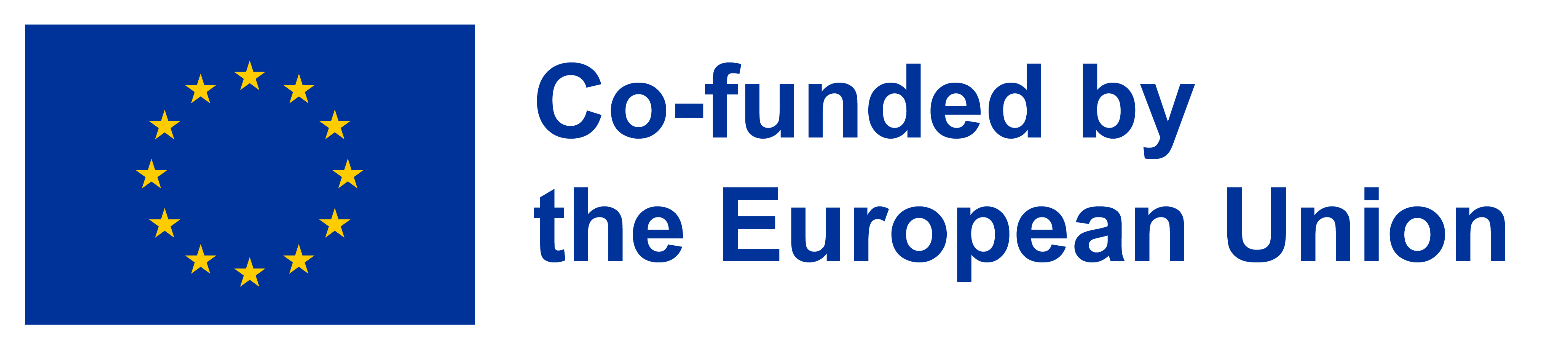}.}

{\footnotesize
%-------------------------------------------------------------------------------------------------------------------------

%------------------------------------------------------------------------------
\end{document}